# Orion+: Automated Problem Diagnosis in Computing Systems by Mining Metric Data

Charitha Saumya

Nomchin Banga

Shreya Inamdar

Purdue University

School of Electrical and Computer Engineering

# Introduction

Nowadays, distributed systems are the necessity of almost all big enterprises. It is a programmer's nightmare to encounter a bug which causes failures in the system and leads to a crash on such a large infrastructure. With the ever increasing code sizes and processing needs, a tool is required that is able to assist a programmer in figuring out potential causes of a bug and minimizing time taken for debugging, hence rectifying it quickly.

This problem has existed since the inception of distributed systems and continues to be a major development and monitoring bottleneck till date. Numerous solutions have been proposed to assist programmers in determining deterministic as well as byzantine failures in these large-scale systems. A few of the well-noted works in the domain utilized breakpoint debugging, replay tools using system logs [3] and quantitative comparison of performance metrics of different runs [4]. One such solution, called Orion [1], implements an automated diagnosis tool in computing systems by mining metric data. The focus of this tool is on manifest-on-metrics bugs where a faulty program execution will result in an abnormal pattern in system metrics.

Our work focused on improving the functionality of Orion[1]. Orion compares the system metrics at various levels, namely, hardware, OS, middleware and application layer. The previous algorithm [2] is able to isolate the code region associated with the change in a particular metric. However, it does not make use of the association information provided by the stack traces of the normal and abnormal runs. We utilized the execution logs of a process, stored in the form of a stack trace, to narrow down the specified buggy code region to a particular sequence of function calls that contain the bug or are most affected by the bug.

Another aspect of our solution focused on increasing the reliability of Orion's results. The existing algorithm [1] used a single normal run for comparison with an abnormal run. However, the system metrics might differ between various normal runs based on the input parameters and runtime environment. We augmented this functionality by enabling Orion to take into account the system metrics from multiple normal runs. Our solution focused on finding the closest normal run that best emulated the metrics of abnormal run before the process diverged abnormally.

---

[1] This is a distinct system from the Orion of [8] which is a system for parallelizing genomic queries.

The first part of our algorithm focused on refining the granularity of Orion by locating the buggy function call sequence. The basic technique included design of a similarity index for determining the amount of correlation between two sequences of function calls, each belonging to normal and abnormal executions of the process. Following this, a dissimilarity score is calculated for each pair of abnormal and normal windows in decreasing order of similarity index. This score captures the dissimilarity in the behavior of the most similar stack traces between abnormal and normal runs.

The second part of the algorithm focuses on better calibration of the results by determining the buggy metric using multiple normal runs. In a distributed system, multiple threads execute simultaneously which results in slight aberrations in the variance of system metrics of each new run. Therefore, considering the run closest to an abnormal run magnifies the dissimilarities better. We use the existing semantics of Orion for calculating correlation between different windows of abnormal and normal runs. Instead of calculating the correlation for different windows of same normal run, the extended algorithm calculates these correlation vectors across all normal runs and helps in identifying the normal with the closest deviation from the abnormal run.

We benchmarked our work against already established bugs in open source software which have been fixed. This enabled us to verify our findings with the actual diagnosis of the bugs by developers. On running Orion's extended functionality against the bug Hadoop-3067, the algorithm successfully listed the buggy function call sequence among the top three results. In the same case, we were also able to see a difference in the ordering of suspected buggy metrics when run with metric data from different normal runs.

To summarize, we have extended the functionality of Orion to be able to locate the buggy function call sequence over the existing algorithm that isolated the code region. Our algorithm is generic to applications and is able to locate the function call sequence which either contains the bug or is most affected by the bug. We have also increased the reliability of the results by including multiple runs for detecting the faulty metric. We successfully demonstrated Orion+ with Hadoop 3067. Orion+ identified the closest normal run to the abnormal run, which gave the faulty metrics in the most relevant order. It further found the function call sequence that was most affected in a given code region, as well as contained the specific method where the bug was present.

The following sections of the report will further expand on the background research that was conducted during the project, details about the existing implementation of Orion, design overview of Orion+, comprehensive approach to the problem at hand and its implementation specifics, experimentation and results. Finally, we conclude with the insights from the project and a brief discussion about the future scope of the project.

# Background

Today's large-scale distributed applications are complex. With the increasing application complexity, finding the root cause of performance problems in an automated manner is a challenging task. Traditionally, application developers and administrators who have domain knowledge follow breakpoint-based debugging or ad-hoc profiling techniques to manually find the root causes of the performance problems. A performance problem in an application can occur due to a variety of root causes. Majority of the root causes relate to an abnormal pattern in the behavior of one or more metrics in the system. For example, due to a bug, an application can abnormally start opening a large number of file handles without closing them until the system runs out of the file-handle limit.

With such a huge system at hand in production, efficient debugging becomes an important aspect of the development process. Processes, when failing in production may have a negative impact on the users and it is imperative to fix them at the earliest. Therefore, this problem is both of theoretical as well as practical importance and a solution to faster and more efficient debugging of distributed applications will increase the reliability and availability of these systems.

An earlier related work in performance diagnosis area was presented in [4] where first using different metrics, a performance state (called a fingerprint) based on quantiles is established. The fingerprint is then statistically compared with previously known fingerprints that resulted in performance issues in past. Another approach is based on using a replay debugging tool called liblog [3]. It logs the execution of deployed application processes and replays them deterministically, faithfully reproducing race conditions and non-deterministic failures, enabling careful offline analysis. Liblog focuses on long-running services that are particularly prone to the slow-developing and non-deterministic, low-probability faults that resist detection during the testing phase. Prior work [7] has also shown that classification of failures is helped if there is application-specific knowledge that can be fed to the classifier. We aid in that effort through our extraction of application-specific parameters.

Our work is built over Orion [1], which was developed by previous members of the Dependable Computing Systems Lab. Orion is an automated problem diagnosis in computing systems by mining metric data. It focuses on finding the root cause of performance problems that are associated with resource-leaks (where resource could be file- handles, memory, connections etc.). Orion pinpoints the metric and a window that is most highly affected by a failure and subsequently finds top-k code regions that are associated with the problem's origin. ORION's algorithm models the application behavior through pairwise correlations of multiple metrics collected across different system layers. When failure occurs, it finds the correlations that deviate from normal behavior.

Since our work is an extension of the existing functionality of Orion, it is only justified to give a detailed overview of the previous algorithm. The details following herewith form the building blocks for our proposed solution. The figure below gives an overview of the algorithm.

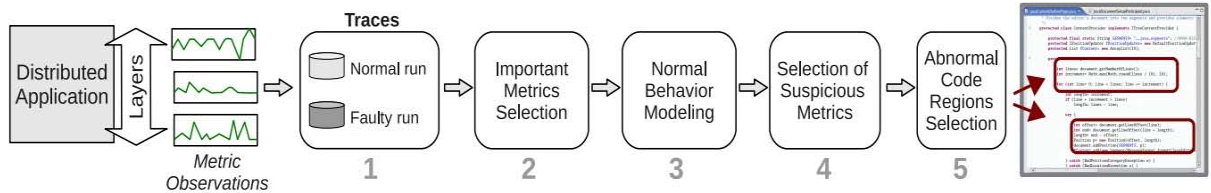

*Figure 1 Problem determination Workflow of Orion[1]*

Orion uses measurements of multiple metrics at different levels in the system, i.e., hardware, OS, middleware and application for problem diagnosis in distributed applications. Given a normal and abnormal trace and log of system metrics, Orion calculates the correlation of pair of metrics between windows split across time. Using this distance matrix, Orion creates a hypersphere of these distance vectors and locates the faulty metric based on the maximum minimum distance between any two normal and abnormal windows.

The second half of Orion, which focuses on identifying the faulty code region, takes in the resulting faulty metric as an input. The algorithm then finds the abnormal windows that contribute most to the anomaly in the buggy metric. The aim of this part of the Orion algorithm is to pinpoint the code regions that might have contributed to the aberration in this particular metric. Therefore, after identifying the most anomalous windows for this faulty metric, Orion create a frequency map for number of unique occurrences of each of the code regions across the most abnormal windows. The code region with the highest are the potential culprits. Exactly which features should be used for the detection and the diagnosis phases is somewhat use case dependent and there is much prior work in feature engineering that is required to develop accurate classifiers [9].

The figure below summarizes the above algorithm in a pictorial format to give more clarity to the reader as to the flow of data, the processing units and the results obtained from the existing functionality of Orion. Each of the lines in the normal and abnormal run windows record the entry and exit of different functions and the values of system metrics at that instant of time. A particular entry 'A/a' symbolizes that the code region is 'A' and the specific method call is 'a'.

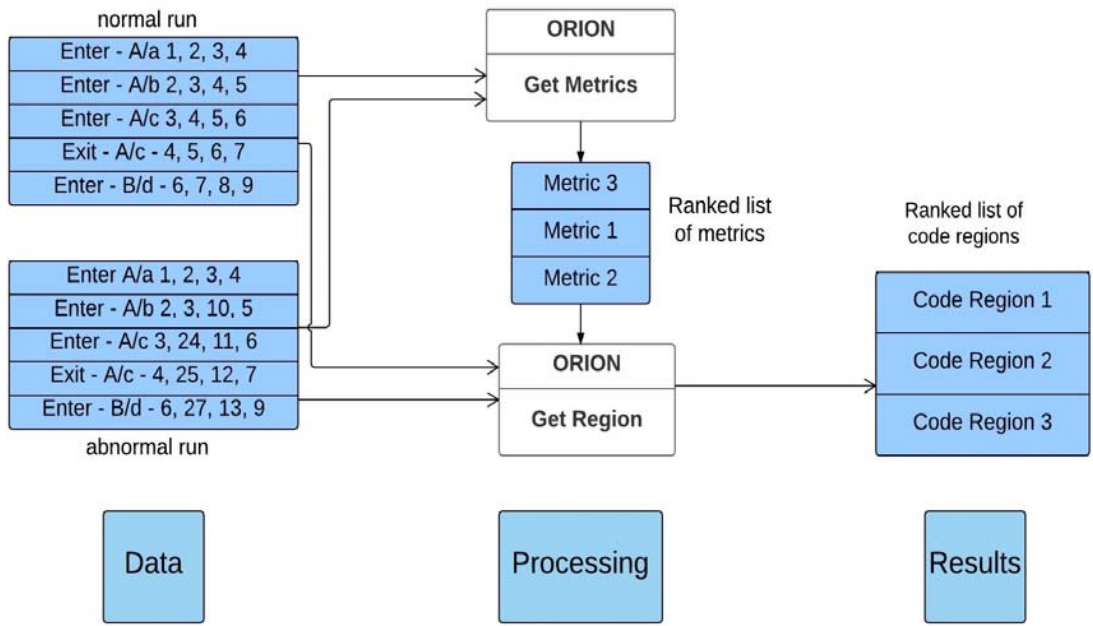

*Figure 2 Orion implementation overview*

# Design Overview

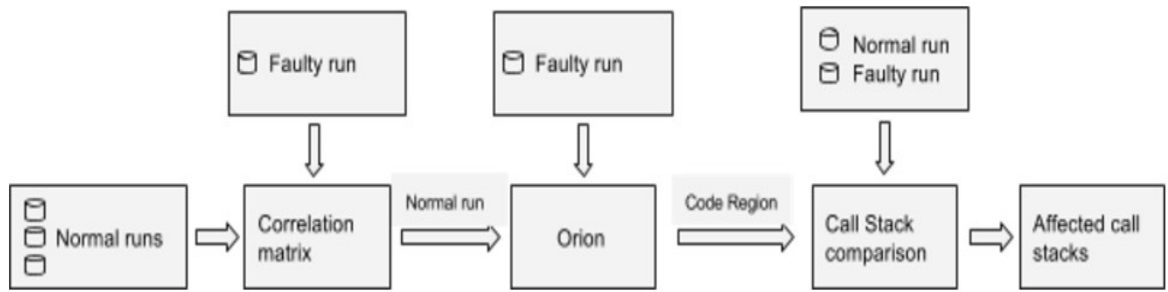

*Figure 3 High level conceptual figure of solution*

To improve the results returned by ORION and to make it more developer friendly, two major design changes were done.

1. Using a most relevant normal run for getting the code regions from ORION
2. Get a call stack from than a code region which is most impacted by the fault

A set of logs of normal runs and an abnormal run are used to collect an accurate normal run by using correlation matrix. The normal run returned by this module is used in Orion to get a code region that has the bug. Based on the logs of normal and abnormal runs call stack is computed for each entry in the logs from the code region returned by Orion. Based on occurrence patterns of similar call stacks, suspicious call stacks are returned to the user.

This design considers call stack changes that result in repetitive blocks, recursive blocks and disjoint blocks in similar call stacks from normal and abnormal runs. For example, repetitive blocks can be expected in case of a fault that results in retries of a method, recursive blocks can be expected in case of a missing termination condition and disjoint blocks in case of an exception handling block in the faulty run. In general similarity for any two call stacks is measured by cardinality of intersection of sets composed of call stack entries. For each specific case, additional length based restrictions are added.

# Solution Details

## Multiple normal runs

The current implementation of Orion takes only one normal run do the statistical comparison with the abnormal run. But in practice the program under test can have multiple executions with completely different statistical profiles. This can be due to differences in initial conditions they are run with. So to have a better understanding of the root cause of the bug/ deviation we need to find a normal run that is closely related to the given abnormal run. To illustrate this idea, consider the following diagram.

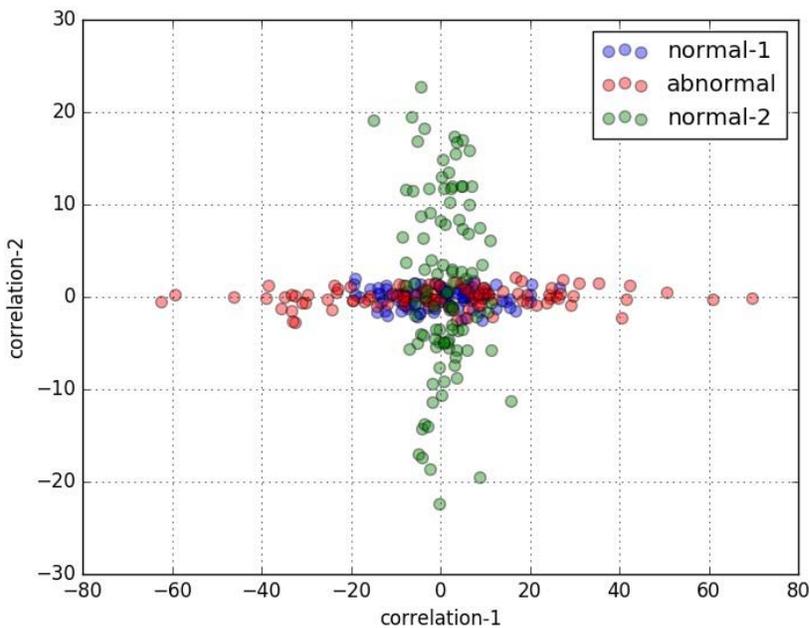

*Figure 4 Overlap of various normal and abnormal runs*

Here x and y directions represent the pair-wise correlation values Orion generates and the plot shows the correlation vector points for 2 normal runs and one abnormal run. For the sake of simplicity, we assume the correlation vectors are 2 dimensional. It is clear in this case that normal-run 1 is closer to the given abnormal run. So comparing normal-run 1 with abnormal run will give better insight on the suspicious metrics. In contrast if we considered normal-run 2 with the abnormal it is highly likely that it will overestimate the deviation and provide incorrect metrics as the suspicious metrics.

**Proposed Algorithm**

Our algorithm to handle this problem is developed on top of the results Orion already generates. What we are interested in is finding the normal run which is closest to the abnormal run in terms of metrics variations.

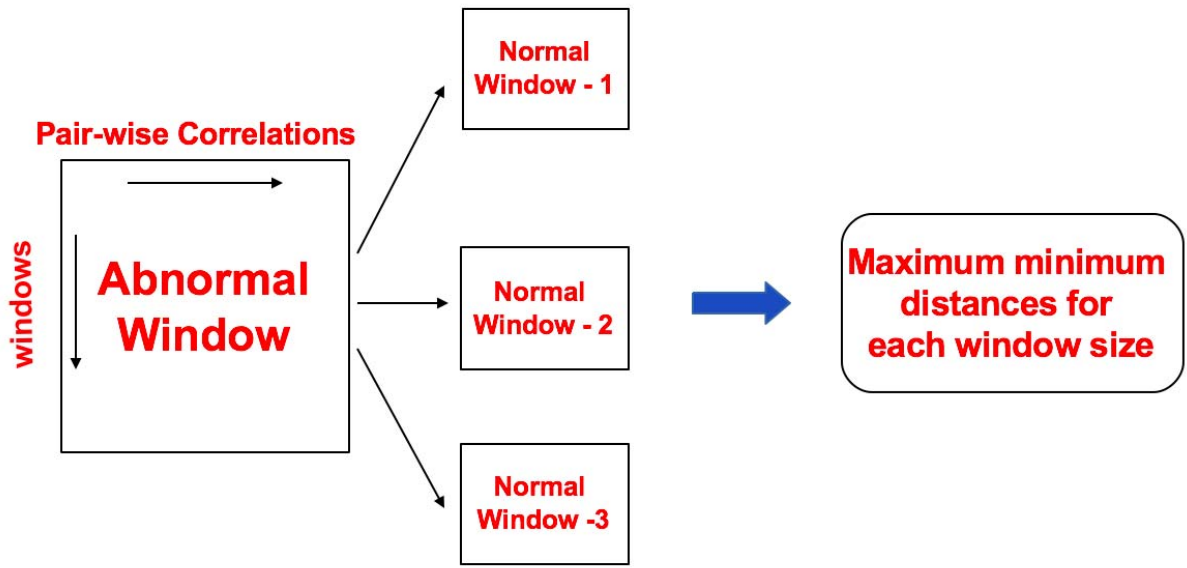

*Figure 5 Flowchart of multiple normal runs*

We take multiple normal runs as input to Orion+ and the abnormal run that we are interested in. We build the algorithm on top of the existing Orion code where the abnormal and normal runs are compared based on their correlation matrix. We extend that relation to calculating correlation between the abnormal run and across all windows of all normal runs. We are interested in the run which is closest to the given abnormal run i.e. most of the windows which are at a maximum minimum distance from the corresponding abnormal run belong to this normal run. We identify this normal run by creating a frequency map of the number of anomalous windows that belong to each normal run. We then find the normal run whose windows contribute the most to the top 25 windows with lowest Euclidean distance from the abnormal windows.

## Finding the faulty stack trace

### Approaches

### LCS

Orion returns the abnormal windows based on the deviation of system metrics of the faulty run from the normal run. Initial assumption was that if a window in abnormal run is farthest from normal run, the code executed in the faulty window and not in normal window will contain functions that execute in the buggy run. Based on this assumption, Longest Common Subsequence algorithm was implemented to get the LCS of log entries across top k anomalous windows, k being the parameter that decides the number of relevant farthest windows. The apparent issue with this approach was the complexity. n-LCS is NP Hard. There were no good approximation algorithms that could be used in our case. Even on implementing this by dynamic programming, it would not be a feasible solution in practical debugging scenarios.

### Graph based approach

To reduce the complexity, a graph based approach was designed. Since LCS inherently imposes a "precedes" relation between every character and all characters following it, we wanted to relax this condition, to reduce the complexity, and honor only the caller-callee relation between entries in log. This was implemented as an edge between a caller and each of its callee functions.

On building the call graph for one out of k abnormal windows, the rest of the call logs were just used to modify the edge weights. For every caller-callee edge in the call stack, if the edge exists in the graph, the edge weight was incremented by 1 and no action otherwise. By this method, the caller-callee edge with maximum weight would be expected to be a path that gets called multiple times in faulty runs and would be the suspicious call sequence. On implementing this, it was observed that the edge weights did not increase as expected. Instead, each edge was either of weight 1 or 2. Even the weaker condition of intersection of log entries across top abnormal windows was empty. On investigating the cause of this behavior turned-out to be the correctness assumption. The metrics change gets accumulated over time and the farthest windows need not necessarily be executing the same lines of code. Thus it was clear that any approach that considered only code across top abnormal windows would not lead to the correct solution. Instead, the entire call log for abnormal run should be considered against entire call log for normal run. The output of ORION for suspicious code regions could still be used.

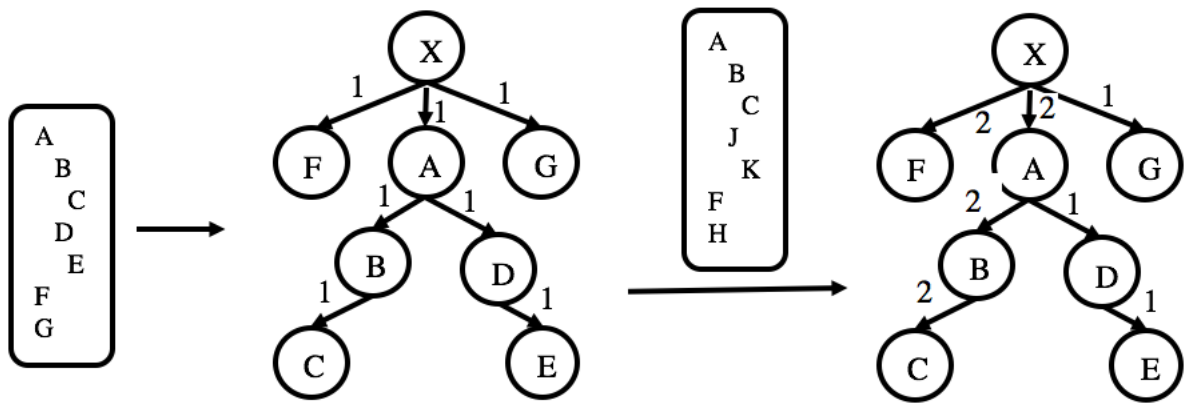

*Figure 6 Graph based approach illustration*

## Stack-based approach

Call stack is collected for each of the methods from the suspicious code region returned by ORION for both normal and abnormal runs. Frequency of each call stack of interest is recorded for normal and abnormal runs. Each call stack from normal run is compared with every call stack from abnormal run based on a distance metric. An equality comparison wouldn't work in our case as there is a possibility of different threads either interleaving the functions in different ways or calling different functions altogether. Distance is measured as a function of number of common functions called in each of the call stacks. This is captured by intersection of the set composed of call stack entries. The frequency of occurrence of each of the calls stacks is recorded. For similar call stacks, if there is a huge difference in frequency, it is noted as a suspicious call stack.

The following patterns are considered while comparing call stacks -

1. Repetitive call stack
2. Recursive call stack
3. Disjoint call stack

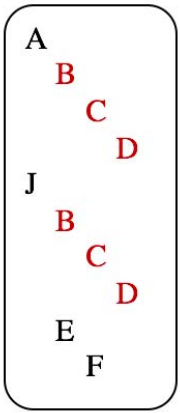
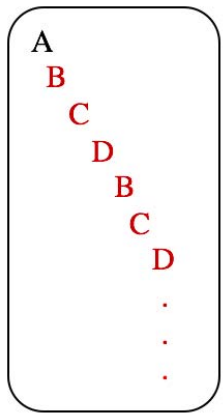
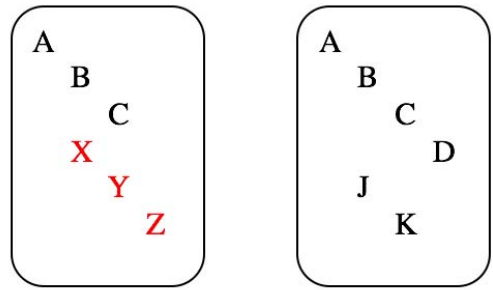

*Figure 7 Illustration of different stack based approach*

**Repetitive call-stack**: When similar code-blocks appear multiple times as compared to the normal run call stack. This shows as a huge difference in frequency of similar call stacks. In this approach, in addition to content of two call stacks, a minimal length difference is rewarded in the similarity metric. This is to remove the cases, where content has a significant overlap just because of having more content in either the abnormal or normal run.

Specifically, initial length constraint is -

$len(abnormal\_stack) - len(normal\_stack) < N$

N is a parameter that needs to be tuned per case basis.

For the call stacks that pass this constraint, the following content criteria needs to be met for similarity.

*len(abnormal_stack int normal stack) = min (len(abnormal_stack), len(normal_stack))*

This criterion imposes that all functions in the normal stack trace must be present in abnormal stack trace. For the functions that pass through, the similarity score is computed as follows -

*Score = freq(abnormal_stack) / freq(normal_stack)*

All call stacks are ranked by their score and the top pair of abnormal and normal stack trace are returned to the developer. Repetitive call stack approach is meant for functions that are called repeatedly in abnormal runs, but this repetition is not captured by the call stack for these functions.

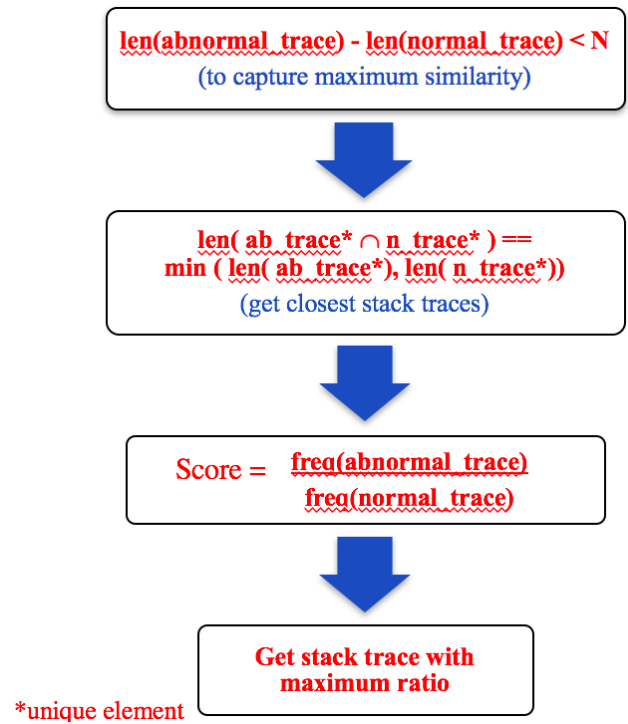

*unique element

*Figure 8 Flowchart of repetitive stack based approach*

**Recursive call-stack** - This approach identifies the code blocks that are repeated, and this repetition is captured in the call stack itself. For such repetitions, the content criteria is more central. Unlike the repetitive call stack approach, this approach rewards longer differences in length when the longer call stacks also have a higher repetition. Based on this, the similarity metric in this case is –

Content:

*len(abnormal_stack int normal_stack) = min(len(abnormal_stack), len(normal_stack))*

This imposes that normal stack trace entries be subset of entries of abnormal stack trace.

Length:

*len(abnormal_stack) - len(normal_stack) > N*

The stack traces for which the condition doesn't meet are taken care of in Repetitive call stack approach. For the pairs of stack traces from normal and abnormal runs that meet the above criteria, score is computed as follows -

*Score = (len(abnormal_stack) - len(normal_stack))/len(abnormal_stack int normal_stack)*

This score rewards the longer abnormal call stacks and those with a smaller normal stack length, when both have considerable overlap. All call stacks are ranked by this score and the top call stack is returned as the most affected call stack in this approach.

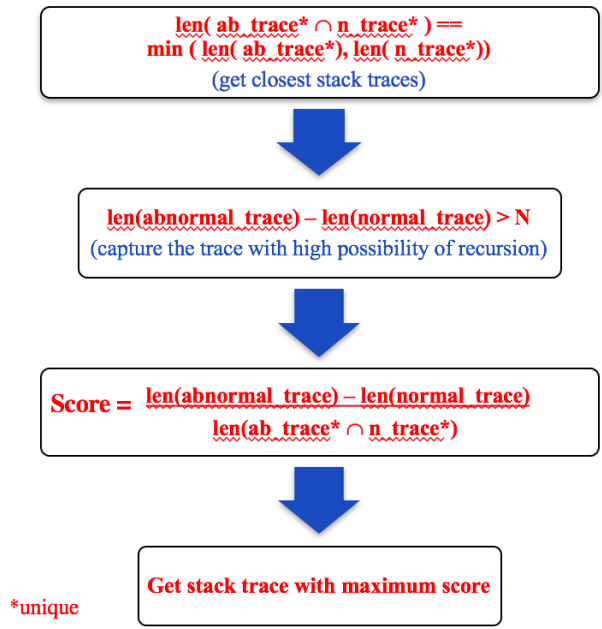

*Figure 9 Flowchart for recursive stack based approach*

**Disjoint call-stack**: In case of a faulty run, a likely scenario would be exception handling/failure handling. In cases like these, a difference code is executed in faulty runs when compared to normal runs and comparing call stacks based on content and length lose their relevance. In this case, just recognizing the code block that was exclusively executed in faulty run is in itself a valuable information that can be used to debug the fault. To account for intersection in parent functions, a parameter rho is used to relax the zero intersection check. The following is the condition to declare that two stack traces are disjoint -

*len(abnormal_stack int normal_stack) < rho*

Thus, every call stack in abnormal run is checked for its presence in normal call stacks' set. Disjoint call stacks are returned as affected code blocks.

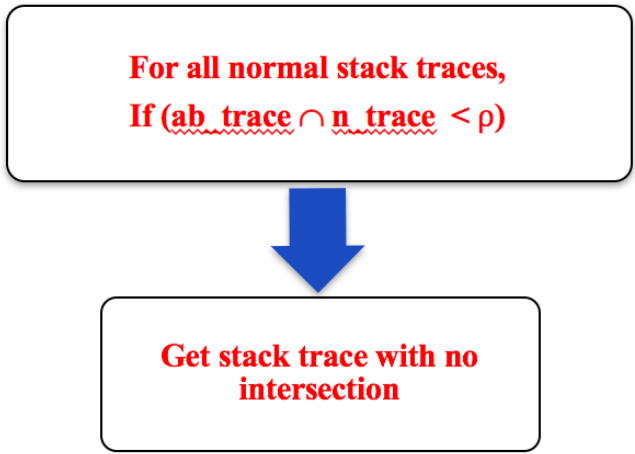

*Figure 10 Flowchart of disjoint stack based approach*

The results from these three approaches are returned to the developer.

# Implementation

Since Orion+ is an extension of Orion, we have extended the functionality in the same language, Python. The program is supported by all versions after Python 2.7. The implementation is platform independent and was tested in Windows, Linux and OS X environments with intel x86 processors.

We modified Orion to input multiple normal runs and output the closest normal run with respect to a particular abnormal run. The code complexity of this implementation is small, () however the time complexity is of the order of $O(n*m*k)$, where n is the number of normal runs, m is the number of windows for a particular split and k defines the number of ways in which the splits are performed (generally a constant).

Our implementation for determining the faulty stack trace focuses on the stack trace of methods belonging to a particular code region deemed faulty by Orion. Since the logs collected by the profiling tool consist of all code regions, we filter the methods pertaining to the potentially faulty code region. This is in resonance with the fact that the buggy function call sequence should belong to the faulty code region while reducing the space and time complexity of the algorithm as well.

The implementation of our working solution is concerned with developing three different use cases for the three distinct ways in which a buggy stack trace exposes itself. The implementation is a simple one in terms of the lines of code that were required to implement it. The complexity of piecewise algorithm is as follows:

1. Creating a dictionary of multiple stack traces, each pertaining to unique methods in the faulty code region - $O(N)$ time complexity, where N is the length of the full logs produced by the profiling metric. $O(I*L*K)$ space complexity, where I is the number of unique methods of faulty code region, L is the average length of each of their stack traces and K is the number of stack traces of a particular function.

2. For getting the relation between abnormal and normal stack traces for repetitive, recursive and disjoint stack traces – O(n*m) time complexity, where n is the number of stack traces in abnormal run and m is the number of stack traces in normal run. Since we consider each stack trace of abnormal run against all stack traces of normal run the complexity is quadratic in nature. O(n*L) space complexity for storing the resultant stack traces in decreasing order of scores, where N is the number of stack traces and L is the maximum length of a particular stack trace.

Hence, the overall time complexity of the algorithm is O(N + n*m) in worst case and space complexity is (I*J*K + n*L). However, these complexities are oblivious of the complexities involved in the functions internal to the algorithm, like intersection of sets, dictionary access among others.

# Experiments and Results

## Experimental setup

We conducted our experiments with Hadoop – 3067 bug. The bug dealt with open sockets that were not being closed by the application. This resulted in an unprecedented increase in the number of open file descriptors. This bug can be reproduced by running a test case, specifically TestCrcCorruption of Hadoop's DFSClient module.

We cloned the Hadoop version where the bug had surfaced, i.e. version 17.0. However, the bug had been patched in the same version due to which the buggy version was not available. We forked the unpatched version from Apache Hadoop's website and used it to run our tests. We setup the Hadoop system on our personal system and ran the unpatched version with this test as well as the patched version to reproduce normal and abnormal runs.

Further, we dived into SystemTap [5] and Javassist [6] to generate the system metrics. However, SystemTap required specific environment to run in and Javassist only provided stack trace as an output. We then explored the profiling tools [7] developed by one of the members of DCS lab, Ignacio Laguna to generate the system metrics that matched the metrics generated for Orion earlier. We the used this process profiling tool to generate metrics at each ENTRY and EXIT point of the application during its run, both for normal and abnormal runs. The 17.0 version of Hadoop uses ANT and therefore we had to tweak the build.xml file to Hadoop's unit test case with process profiler.

## Multiple Normal Run Experiment

Using the above setup, we generated three normal runs for Hadoop's testCrcCorruption test case with different inputs, namely, io.bytes.per.checksum and dfs.block.size. On passing these normal runs to Orion, we get the below mentioned graph for the three normal runs. The x axis marks the different windows, with each of the three colored bars belonging to different normal runs. The y axis denotes the Euclidean distance between the corresponding normal and abnormal windows. The highest bars denote the farthest of the three runs.

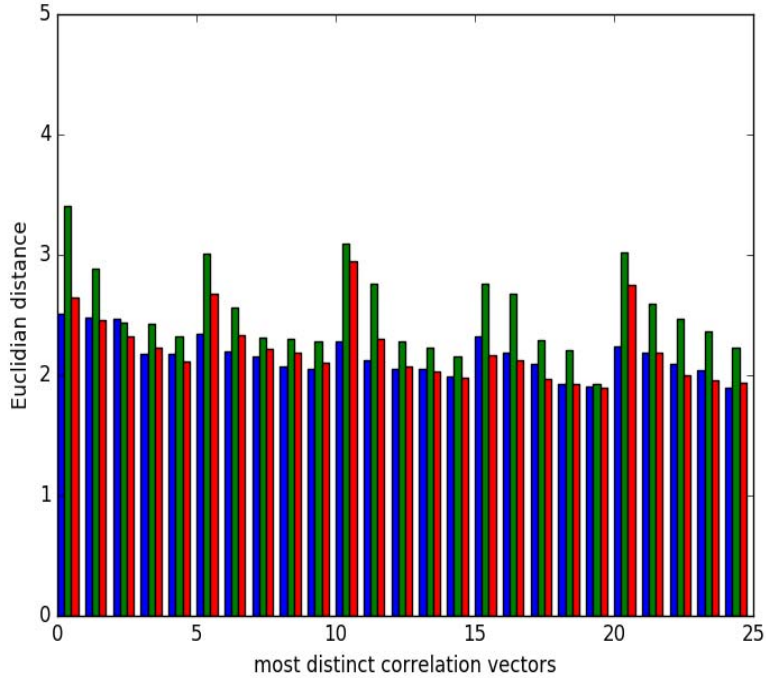

*Figure 11 Graphical representation of correlation between abnormal and normal runs for different windows*

As we can see from above graph, the normal run whose windows are denoted by blue bars is the closest run to the abnormal matrix. The algorithm proposed by us works in a similar way. We first calculate the correlation between each pair of normal and abnormal window. Following which, we sort these windows in increasing order of correlation. The run which contributes the most to the top 25 windows of this sorted list is taken as the closest window to the specific abnormal run.

As can be seen more clearly in the graph below, the blue run contributes the majority of windows in the top 25 windows that are at maximally minimum distance from the abnormal run, Orion outputs the blue run as the closest normal run.

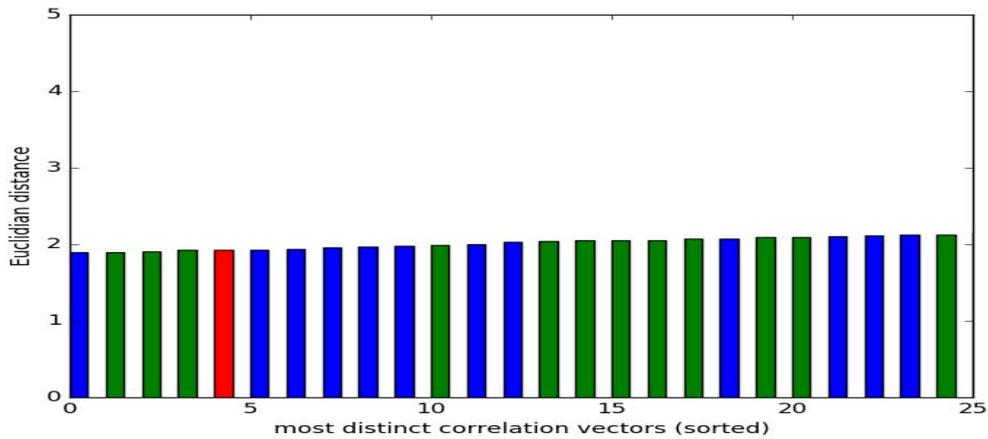

We further experimented with two normal runs by passing them into previous algorithm of Orion and comparing the top 10 metrics that are output as the potential buggy metrics.

```
========== Top-3 Abnormal Metrics ==========
Format: [Rank] [Metric]
[1]: vsize
[2]: rchar
[3]: read_bytes
========== Other Metrics ==========
[4]: minflt
[5]: utime
[6]: num_threads
[7]: rss
[8]: stime
[9]: wchar
[10]: num_file_desc
chartha@chartha-inspiron-7559:~/GIT/orion$
```

*Figure 12 Top 10 abnormal metrics with normal run 1*

```
========== Top-3 Abnormal Metrics ==========
Format: [Rank] [Metric]
[1]: vsize
[2]: stime
[3]: minflt

========== Other Metrics ==========
[4]: rss
[5]: utime
[6]: write_bytes
[7]: num_threads
[8]: rchar
[9]: wchar
[10]: num_file_desc
chartha@chartha-inspiron-7559:~/GIT/orion$
```
*Figure 13 Top 10 abnormal metrics with normal run 2*

As can be seen from the two figures above, Orion outputs different ranking of top ten metrics for the two corresponding normal runs.

This experiment validated our algorithm that different normal runs may differ in the differences in their metrics depending on the input conditions and environment in which they are run. It also validated that Orion+ does return the closest normal run to the corresponding abnormal run.

**Faulty Function Sequence Experiment**

We also conducted experiment for validating our algorithm for fining the faulty function sequence using the same experimental setup. We used the normal run output by the above experiment to produce the faulty metric. The faulty metric was num_file_desc which was in accordance with the earlier results of Orion.

```
========== Top-3 Abnormal Metrics ==========
Format: [Rank] [Metric]
[1]: rss
[2]: num_file_desc
[3]: minflt

========== Other Metrics ==========
[4]: write_bytes
[5]: num_threads
[6]: vsize
[7]: wchar
[8]: stime
[9]: rchar
[10]: utime
```

Figure 14 Top 10 abnormal metrics with closest normal run

Using this metric, we generated the faulty code regions from a second pass of Orion and using this code region, tried to narrow down the scope of the code region to the buggy function call sequence.

```
$ ./orion.py -n data/hadoop_case/normal_med.dat -a data/hadoop_case/abnormal_med.dat --select-regions -m num_file_desc
Localization module loaded.
Correlation analysis module loaded.
[ORION]: Normal File: data/hadoop_case/normal_med.dat
[ORION]: Abnormal File: data/hadoop_case/abnormal_med.dat
[ORION]: Finding outliers...

========== Top-3 Abnormal Code Regions ==========
[1]:
    [218] org/apache/hadoop/dfs/DFSClient
[2]:
    [183] org/apache/hadoop/dfs/BlocksMap
[3]:
    [172] org/apache/hadoop/dfs/DataNode
```

Figure 15 Result for buggy code regions for num_file_desc

We ran Orion+ with the class name as DFSClient which was the buggy code region output by Orion. Orion+ successfully returned the below figure as a result to the input stack traces for normal and abnormal runs.

```
org/apache/hadoop/dfs/DFSClient$DFSOutputStream$DataStreamer$run
org/apache/hadoop/dfs/DFSClient$DFSOutputStream$writeChunk
org/apache/hadoop/dfs/DFSClient$access$500
org/apache/hadoop/dfs/DFSClient$checkOpen
org/apache/hadoop/dfs/DFSClient$DFSOutputStream$close
org/apache/hadoop/dfs/DFSClient$DFSOutputStream$closeInternal
org/apache/hadoop/dfs/DFSClient$DFSOutputStream$flushInternal
org/apache/hadoop/dfs/DFSClient$DFSOutputStream$access$1700
org/apache/hadoop/dfs/DFSClient$DFSOutputStream$ResponseProcessor$run
org/apache/hadoop/dfs/DFSClient$DFSOutputStream$access$1600
org/apache/hadoop/dfs/DFSClient$DFSOutputStream$closeThreads
org/apache/hadoop/dfs/DFSClient$DFSOutputStream$DataStreamer$close
org/apache/hadoop/dfs/DFSClient$DFSOutputStream$access$1800
org/apache/hadoop/dfs/DFSClient$DFSOutputStream$isClosed
```
*Figure 16 Orion+ faulty function call sequence*

As can be seen from the above figure, Orion+ returns the stack trace for isClosed() function call of the DFSClient code region. This function was the most affected part of the code because of the bug 3067. This can also be validated from the discussion of JIRA ticket from Apache for this bug (https://issues.apache.org/jira/browse/HADOOP-3067).

# Discussion

The proposed solution addresses common pattern of faulty call stacks - repetitive, recursive and disjoint. However, it is not necessarily exhaustive. Further experiments with different bugs would be required to experimentally conclude a confidence level or range of bugs covered by the three call stack based approaches.

The call stack based approach also takes in parameters for maximum length difference, minimum length difference and maximum allowed intersection for disjoint call stacks. These parameters are highly dependent on the bug at hand and need to be tuned based on the level of sensitivity to faulty runs of code. With standard assumptions on sensitivity, the parameter tuning could be automated based on the call log.

# Conclusion

This work presents the suspicious code at a finer granularity of call stack rather than code region, which was being returned by Orion. Call stack based comparison returns call stacks that are most impacted by the bug and save developer time to debug from scratch. This solution has polynomial complexity and hence can be implemented practically.

To improve this work further, current comparison of call stacks in normal and abnormal runs which is purely based on content and length can be improved. If a phase-stamp could be added to the call stacks, based on which phase the code is executing in, the accuracy can be improved by ensuring that appropriate call stacks are only compared, as well as the space and time complexity can be made better by limited comparisons.